\documentclass[fleqn,usenatbib]{mnras}

\usepackage{newtxtext,newtxmath}
\usepackage[T1]{fontenc}

\DeclareRobustCommand{\VAN}[3]{#2}
\let\VANthebibliography\thebibliography
\def\thebibliography{\DeclareRobustCommand{\VAN}[3]{##3}\VANthebibliography}

\usepackage[dvipsnames]{xcolor}
\usepackage{graphicx}	
\usepackage{amsmath}	
\usepackage{orcidlink}

\title[Fine-structure transitions of Si with H and S with H]{Fine-structure transitions of Si and S induced by collisions with atomic hydrogen}

\author[P. G. Yan and J. F. Babb]{
Pei-Gen Yan\orcidlink{0000-0003-1623-1391}\thanks{E-mail: peigen.yan@cfa.harvard.edu}
and 
James F. Babb\orcidlink{0000-0002-3883-9501}\thanks{E-mail: jbabb@cfa.harvard.edu}
\\
Center for Astrophysics \textbar\ Harvard \& Smithsonian, MS 14, 60 Garden St., Cambridge, MA 02138
}

\pubyear{2023}
\begin{document}
\label{firstpage}
\pagerange{\pageref{firstpage}--\pageref{lastpage}}
\maketitle

\begin{abstract}
Using a quantum-mechanical close-coupling method, we calculate cross sections for fine structure excitation and relaxation of Si and S atoms in collisions with atomic hydrogen.
Rate coefficients are calculated over a range of temperatures for astrophysical applications.
We determine the temperature-dependent critical densities for the relaxation of Si and S in collisions with H and compare these to the critical densities for collisions with electrons.
The present calculation should be useful in modeling  environments exhibiting the [\ion{S}{i}] 25~{\textmu}m and [\ion{S}{i}] 57~{\textmu}m far-infrared emission lines or where cooling  of S and Si by collisions with H are of interest.
\end{abstract}


\section{Introduction} \label{sec:intro}

The abundances of elements are particularly important parameters for understanding the composition of the gas and dust in the interstellar medium (ISM) and the physics and chemistry behind it.
Silicon and sulfur are abundant elements that are arguably less-studied than, respectively, isovalent carbon and oxygen. 
Collisional rates for cooling of Si and S in collisions with H, such as we provide here, are useful~\citep{Hollenbach1989} and may be efficiently calculated and presented together, similarly to complementary calculations that were given for C and O in collisions with H, e.g.~\citet{lr77,A07}.
In the case of sulfur, observed concentrations in dense regions of the interstellar medium (ISM) appear significantly less than those in diffuse clouds~\citep{Savage1998,Joseph1986}.
Observations of sulfur fine-structure lines can assist in trying to fill this ``missing sulfur'' gap.
For example, the [\ion{S}{i}] 25~{\textmu}m emission line was observed in shocked environments by \citet{Neufeld2009,Goicoechea2012,Anderson2013} using the \textit{Spitzer} Infrared Spectrograph (IRS) and by \citet{Rosenthal2000} with the Short-Wavelength-Spectrometer on the Infrared Space Observatory (ISO-SWS). 
In another application, recently,  Betelgeuse was observed using the Echelon Cross Echelle Spectrograph (EXES) on SOFIA, where the [\ion{S}{i}] 25~{\textmu}m line
was used to investigate circumstellar flow~\citep{Harper2020}. 
Since emission from the sulfur atom can arise in these certain circumstellar 
(wind of Betelgeuse)
and neutral shocked (C-shock) environments 
of the ISM, it is useful to have predictions
of the critical densities (the densities where 
collisional deexcitation rates equal the
radiative decay rates) for collisions with neutral hydrogen
to reliably establish the diagnostic potential of sulfur fine-structure lines
at various densities and temperatures 
in such environments where warm atomic hydrogen is present.

In the present paper, we present our calculations on the fine-structure transitions of Si and S induced by collisions with atomic hydrogen,
\begin{equation}\label{OIIIH}
\textrm{A}(^3P_j)+ \textrm{H}(^2S_{1/2})\rightarrow \textrm{A}(^3P_{j'})+ 
\textrm{H}(^2S_{1/2})\,,
\end{equation}
where A can be Si or S and $j$ and $j'$ can be $0$, $1$ or $2$, respectively.
In collisions with H, there is little theoretical work on the Si and S systems, compared to C and O where fully quantum calculations are available. 
Semiclassically calculated rate coefficients were given by \citet[Eq.~(42)]{Bahcall1968}, \citet[Table 4]{Tielens1985}, and by \citet[Table 8]{Hollenbach1989}.
The present rate coefficients obtained from fully-quantum calculated cross sections could be used for modeling or simulation and other astrophysical applications. 

\begin{figure}
	\includegraphics[width=\columnwidth]{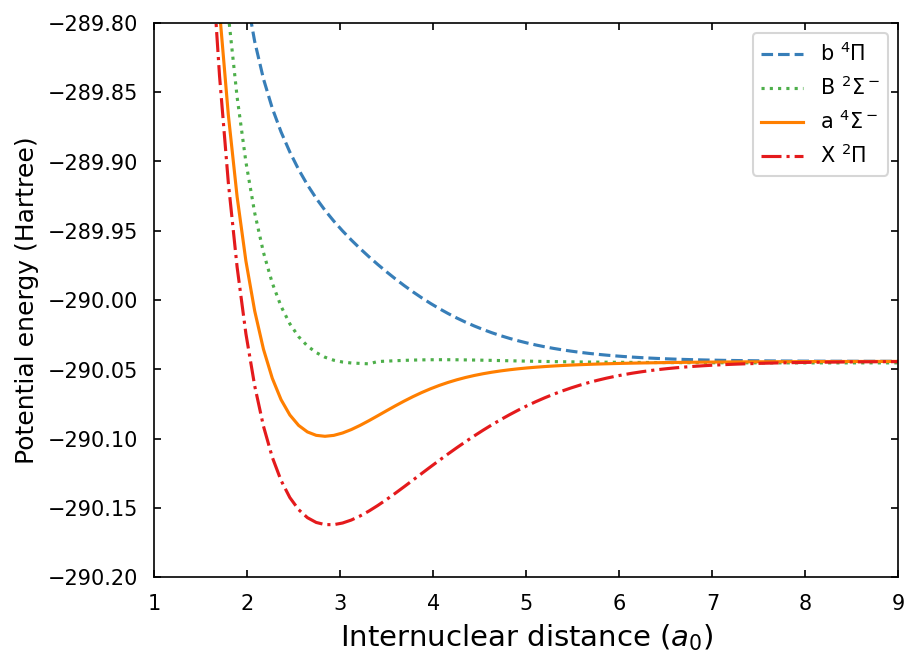}
    \caption{Potential energies for the SiH system as functions of the internuclear distance $R$,
    labeled from the top down: b$^4\Pi$ (blue dashed line), B$^2\Sigma^-$ (green dotted line) , a$^4\Sigma^-$ (orange solid line),   X$^2\Pi$ (red dashdot line).  }
    \label{fig:SiHpot}
\end{figure}

\begin{figure}
	\includegraphics[width=\columnwidth]{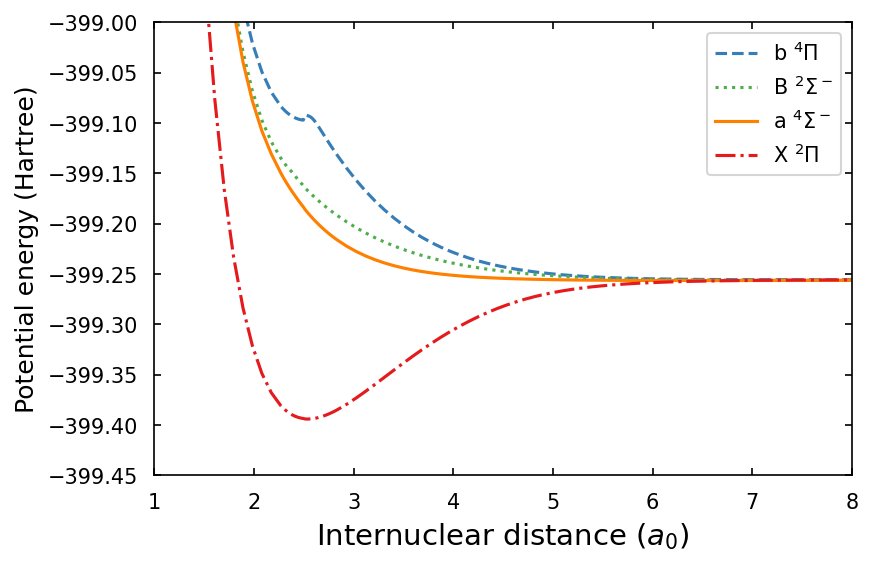}
    \caption{Potential energies for the SH system as functions of the internuclear distance $R$,
    labeled from the top down: b$^4\Pi$ (blue dashed line), B$^2\Sigma^-$ (green dotted line) , a$^4\Sigma^-$ (orange solid line),   X$^2\Pi$ (red dashdot line). }
    \label{fig:SHpot}
\end{figure}

\section{Theoretical Models}\label{TM}
The quantum-mechanical application of
scattering theory
to fine-structure changing collisions
has a long history, see, for example, \citet{lr77,LR77_1,flower_molcol:_2000,Roueff90}
and references therein.
Our approach is 
detailed in \citet{yan_babb_fine-structure_2022,yanbabb2023}.
Here, we give a brief summary to provide specific information for the present calculations. 
The fine-structure splitting energies $\varepsilon_{j_a}$ are adopted from NIST~\citep{NIST_ASD}, where for Si (${j_a}=0,1,2$)
\begin{equation}\label{FinE_Si}
\varepsilon_{0}=0, \qquad \varepsilon_{1}=77.115\;\mathrm{cm}^{-1},\qquad \varepsilon_{2}=223.157\;\mathrm{cm}^{-1} \,,
\end{equation} 
and 
for S (${j_a}=2,1,0$) 
\begin{equation}\label{FinE_S}
\varepsilon_2=0, \qquad \varepsilon_1=396.055\;\mathrm{cm}^{-1}, \qquad \varepsilon_0=573.640\; \mathrm{cm}^{-1} \,.
\end{equation} 

The cross sections for fine-structure transitions are given by
\begin{equation} \label{cross}
\sigma_{j_a \rightarrow j'_a} (E)=\sum_{J} \frac{\pi}{k_{j_a}^2}\frac{2J+1}{(2j_a+1)(2j_b+1)}\sum_{j_{ab}lj'_{ab}l'}|T^J_{j'_aj'_{ab}l';j_aj_{ab}l}|^2\,, 
\end{equation}
where $\sigma_{j_a \rightarrow j'_a}^J(E)$ are the partial cross sections, $k_{j_a}$ is  the wave number defined by $k^2_{j_a}=2\mu(E-\varepsilon_{j_a})$, $\mu$ is the reduced mass of systems $a$ and $b$, $E$ is the total collision energy, and $\varepsilon_{j_a}$ is the fine-structure state splitting energy of the Si or S atom.
The $T$ matrix is defined by $T^J=-2i K^J(I-iK^J)^{-1}$, where $K^J$ is the open channel reaction matrix defined in \citet{johnson_multichannel_1973}.
To solve for the radial wave functions, which determine the scattering matrices, we used the quantum close-coupling formalism with the wave functions expanded in the space-fixed basis~\citep{yan_babb_fine-structure_2022,yanbabb2023}. 

\begin{table}
	\centering
	\caption{Rate coefficients (in units of 10$^{-9}$ cm$^3$s$^{-1}$) for the fine-structure excitation  and relaxation of Si($^3P_j$) by H.}
	\label{tab:sihrate}
	\begin{tabular}{lccccccccr} 
		\hline
         \multicolumn{1}{c}{$T$(K)} 
		  & $j=0 \rightarrow j'=1$ &  $j=0 \rightarrow j'=2$ &  $j=1 \rightarrow j'=2$\\
		\hline
		100 & 0.280 & 0.017 & 0.076\\
		200 & 0.556 & 0.122 & 0.260\\
		500 & 0.917 & 0.493 & 0.657\\
		700 & 1.037 & 0.695 & 0.843\\
		1000 & 1.163 & 0.943 & 1.067\\
		2000 & 1.433 & 1.505 & 1.580\\
		5000 & 1.867 & 2.395 & 2.440\\
		7000 & 2.084 & 2.793 & 2.838\\
		10000 & 2.374 & 3.271 & 3.323\\
		\hline
		\multicolumn{1}{c}{$T$(K)} & $j=1 \rightarrow j'=0$ &  $j=2 \rightarrow j'=0$ & $ j=2 \rightarrow j'=1$\\
		\hline
	    10 & 0.238 & 0.043 & 0.248\\
	    20 & 0.247 & 0.044 & 0.272\\
	    50 & 0.255 & 0.058 & 0.319\\
    	70 & 0.265 & 0.069  & 0.342\\
		100 & 0.280 & 0.084 & 0.370\\
		200 & 0.322 & 0.120 & 0.443\\
		500 & 0.381 & 0.186 & 0.598\\
		700 & 0.405 & 0.219 & 0.682\\
		1000 & 0.433 & 0.259 & 0.788\\
		2000 & 0.505 & 0.353 & 1.052\\
		5000 & 0.636 & 0.510 & 1.526\\
		7000 & 0.705 & 0.584 & 1.754\\
		10000 & 0.800 & 0.675 & 2.036\\
		\hline
	\end{tabular}
\end{table}

\begin{table}
	\centering
	\caption{Rate coefficients (in units of 10$^{-9}$ cm$^3$s$^{-1}$) for the fine-structure excitation  and relaxation of S($^3P_j$) by H.}
	\label{tab:shrate}
	\begin{tabular}{lccccccccr} 
		\hline
         \multicolumn{1}{c}{$T$(K)} 
		  & $j=1 \rightarrow j'=0$ &  $j=2 \rightarrow j'=0$ &  $j=2 \rightarrow j'=1$\\
		\hline
		100 & 0.003 &  & \\
		200 & 0.023 &  & 0.008\\
		500 & 0.091 & 0.006 & 0.076\\
		700 & 0.126 & 0.013 & 0.130\\
		1000 & 0.167 & 0.025 & 0.207\\
		2000 & 0.250 & 0.070 & 0.420\\
		5000 & 0.369 & 0.178 & 0.813\\
		7000 & 0.422 & 0.235 & 0.991\\
		10000 & 0.492 & 0.304 & 1.204\\
		\hline
		\multicolumn{1}{c}{$T$(K)} & $j=0 \rightarrow j'=1$ &  $j=0 \rightarrow j'=2$ & $ j=1 \rightarrow j'=2$\\
		\hline
	    10 & 0.224 & 0.020 & 0.149\\
	    20 & 0.212 & 0.022 & 0.154\\
	    50 & 0.216 & 0.029 & 0.174\\
    	70 & 0.227 & 0.035  & 0.189\\
		100 & 0.248 & 0.045 & 0.208\\
		200 & 0.319 & 0.075 & 0.262\\
		500 & 0.487 & 0.160 & 0.410\\
		700 & 0.569 & 0.213 & 0.500\\
		1000 & 0.665 & 0.290 & 0.621\\
		2000 & 0.870 & 0.531 & 0.942\\
		5000 & 1.194 & 1.080 & 1.557\\
		7000 & 1.351 & 1.355 & 1.841\\
		10000 & 1.554 & 1.693 & 2.185\\
		\hline
	\end{tabular}
\end{table}

\begin{figure}
	\includegraphics[width=\columnwidth]{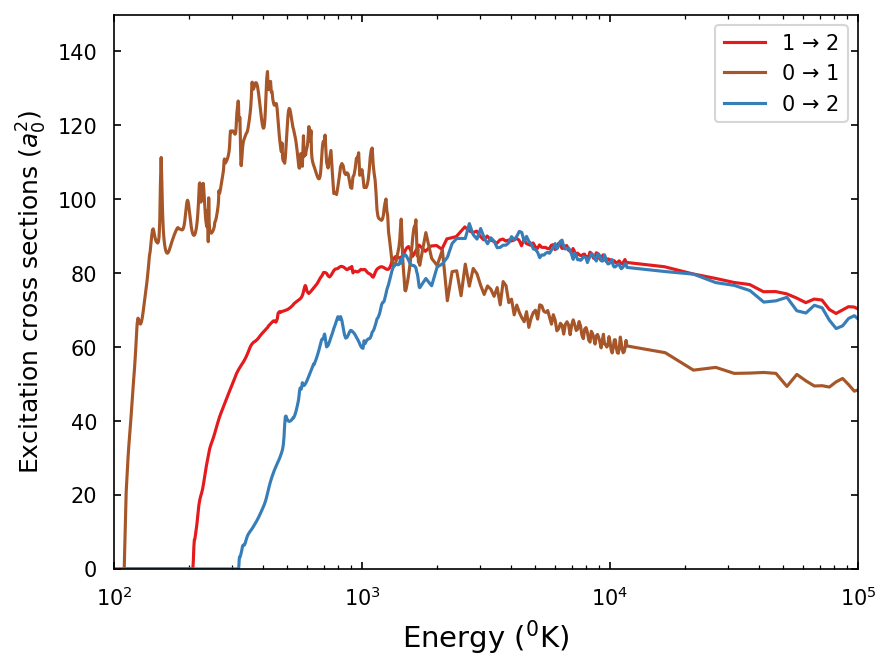}
	\includegraphics[width=\columnwidth]{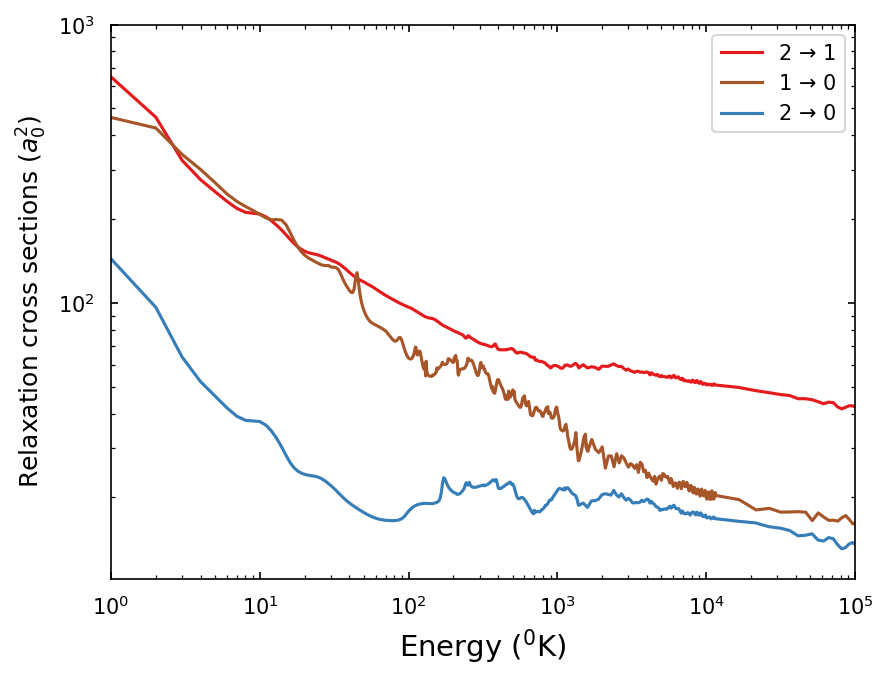}
    \caption{Excitation and relaxation cross sections for fine-structure transitions of Si in collision with H. The three transitions $(j\rightarrow j')$ are labeled with different colors. }
    \label{fig:SiExsct}
\end{figure}

\begin{figure}
	\includegraphics[width=\columnwidth]{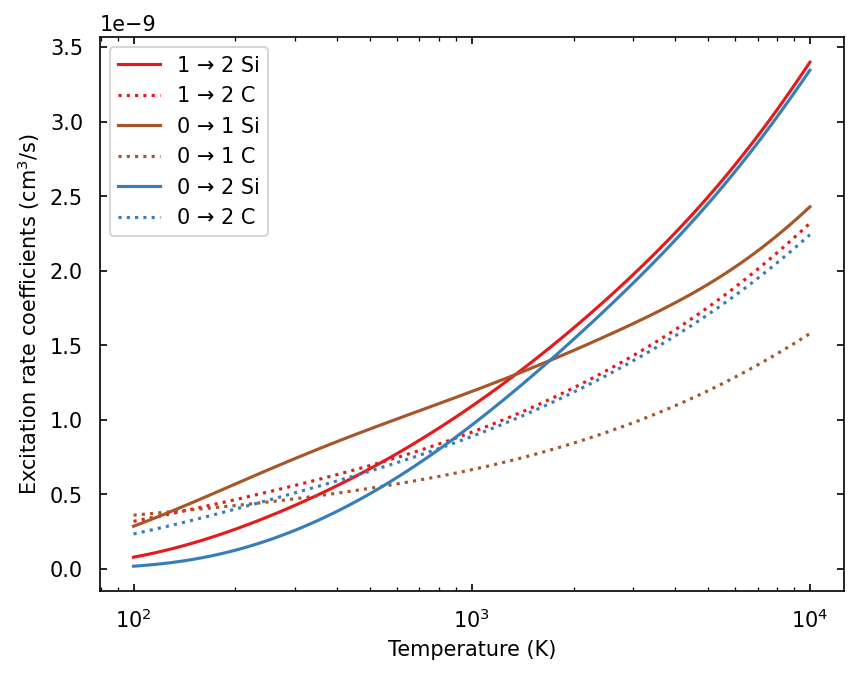}
	\includegraphics[width=\columnwidth]{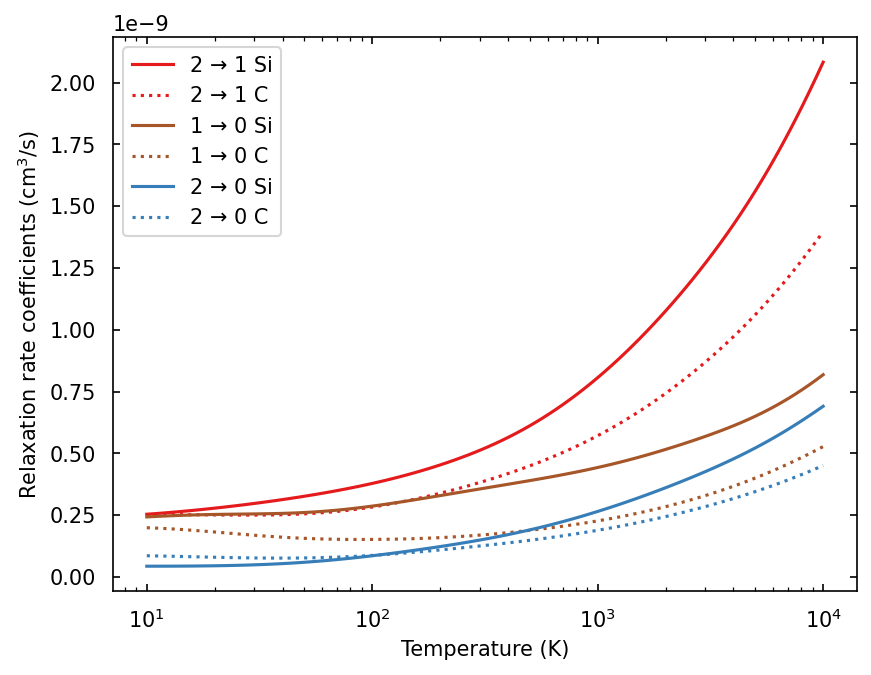}
    \caption{Excitation and relaxation rate coefficients for fine-structure transitions of Si in collision with H (solid lines),
    present work, and for C in collision with H (dotted lines) from \citet{yanbabb2023}. The three transitions $(j\rightarrow j')$ are labeled with different colors. }
    \label{fig:SiExrate}
\end{figure}

\section{Interatomic Potentials}\label{pot}

For both $\textrm{Si}(^3P)$ and $\textrm{S}(^3P)$ 
the electronic states involved in the fine-structure transitions
with $\textrm{H}(^2S)$ are the
b$^4\Pi$, B$^2\Sigma^-$, a$^4\Sigma^-$, and X$^2\Pi$ states.
The potentials for SiH and SH were obtained by using the multireference configuration interaction Douglas-Kroll-Hess (MRCI-DKH) method with the augmented-correlation-consistent polarized valence 5-tuple zeta (\texttt{aug-cc-pV5Z-dk}) basis within \textsc{Molpro} 2010.1~\citep{MOLPRO_brief} in the $C_{2v}$ Abelian symmetry point group. 
For both SiH and SH these potentials were calculated for internuclear
distances $R$ from $1.13$ to $9.07\,a_0$ at an interval
of $0.095\,a_0$ and from $9.07$ to $10.4\,a_0$ at an interval of $0.19\,a_0$.
Note that we have given the values of $R$ in units of $a_0$,
because the scattering calculations were carried out in atomic units (a.u.) 
($1\,a_0 \approx 0.529\,$\AA{}).

For the SiH system, 
10 molecular orbitals (MOs) are used for the active space: 6 $a_1$, 2 $b_1$ and 2 $b_2$ symmetry MOs. The remaining five electrons are put in the closed-shell orbitals. 
The potentials calculated from \textsc{Molpro}  are shown in Fig.~\ref{fig:SiHpot}. 
Our calculated potentials are in good agreement with~\citet{Zhang2018};
for example, for the X$^2\Pi$ state we find $D_e=3.210\,\textrm{eV}$ at $R_e=1.524\,$\AA{} compared to their value
of $D_e=3.193\,\textrm{eV}$ at  $R_e=1.517\,$\AA{}.

For the SH system, 12 molecular orbitals (MOs) are used for the active space: 6 $a_1$, 3 $b_1$ and 3 $b_2$ symmetry MOs. The remaining five electrons are put in the closed-shell orbitals. 
The potentials calculated from \textsc{Molpro}  are shown in Fig.~\ref{fig:SHpot}.  
Our calculated potentials are in accordance
with those of \citet{Hirst1982}. For the X$^2\Pi$ state, our value of $D_e = 3.77\,\textrm{eV}$ 
agrees  with the value $3.791\,\textrm{eV}$ 
adopted by~\citet{Gorman2019,Csaszar2003}.

In order to have the potentials correlate asymptotically
to the separated atom limits, which are taken to be the reference
energies for the scattering calculations,
the potentials for SiH and SH
as shown in Figs.~\ref{fig:SiHpot} and \ref{fig:SHpot} were shifted in energy and joined at $R=9.83\,a_0$ to the
long-range form $-C_6/R^6$.
Smooth fits were obtained using estimates
of $C_6$ from \citep{Gould2016}; the
values used were $C_6=44.1$~a.u. between $\textrm{Si}(^3P)$ and $\textrm{H}(^2S)$ and 
$C_6=30.1$~a.u. between
 $\textrm{S}(^3P)$ and $\textrm{H}(^2S)$.

\begin{figure}
	\includegraphics[width=\columnwidth]{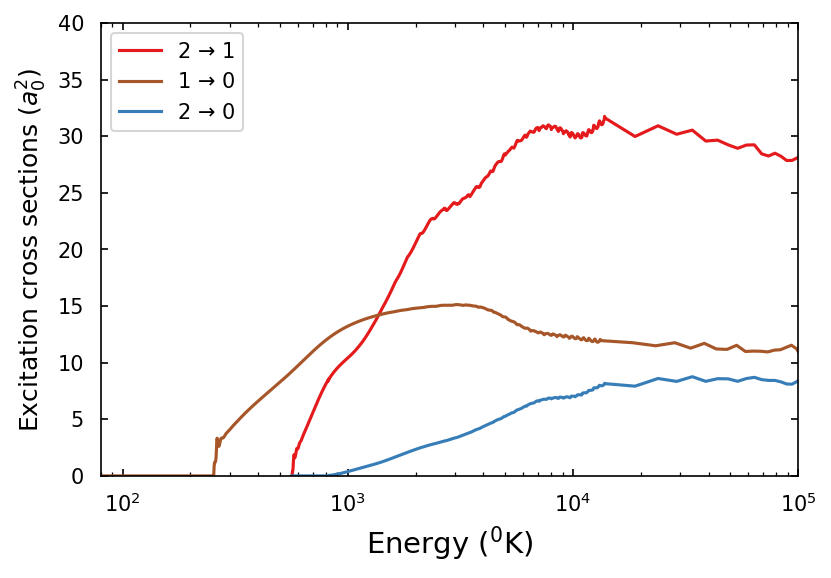}
	\includegraphics[width=\columnwidth]{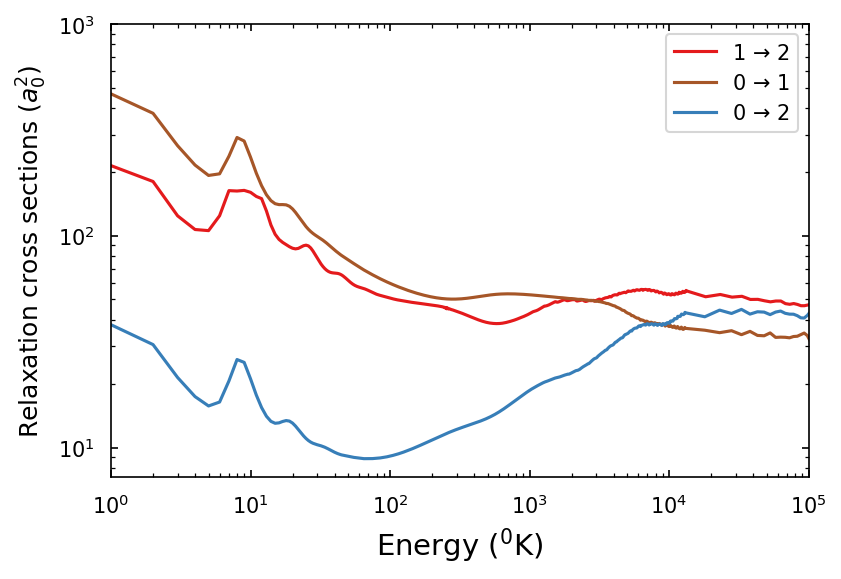}
    \caption{Excitation and relaxation cross sections for fine-structure transitions of S in collision with H. The three transitions $(j\rightarrow j')$ are labeled with different colors.}
    \label{fig:SExsct}
\end{figure}

\begin{figure}
	\includegraphics[width=\columnwidth]{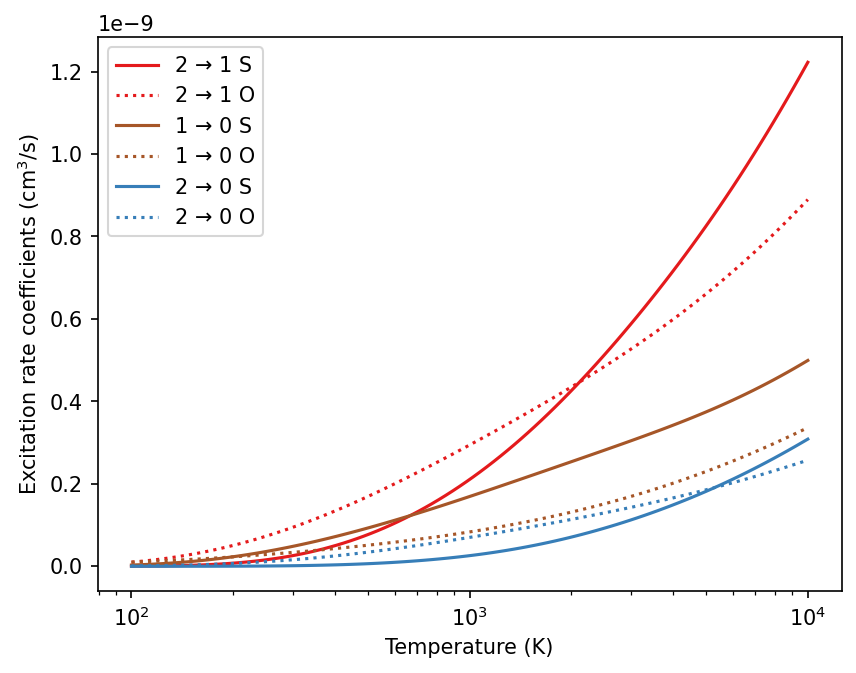}	\includegraphics[width=\columnwidth]{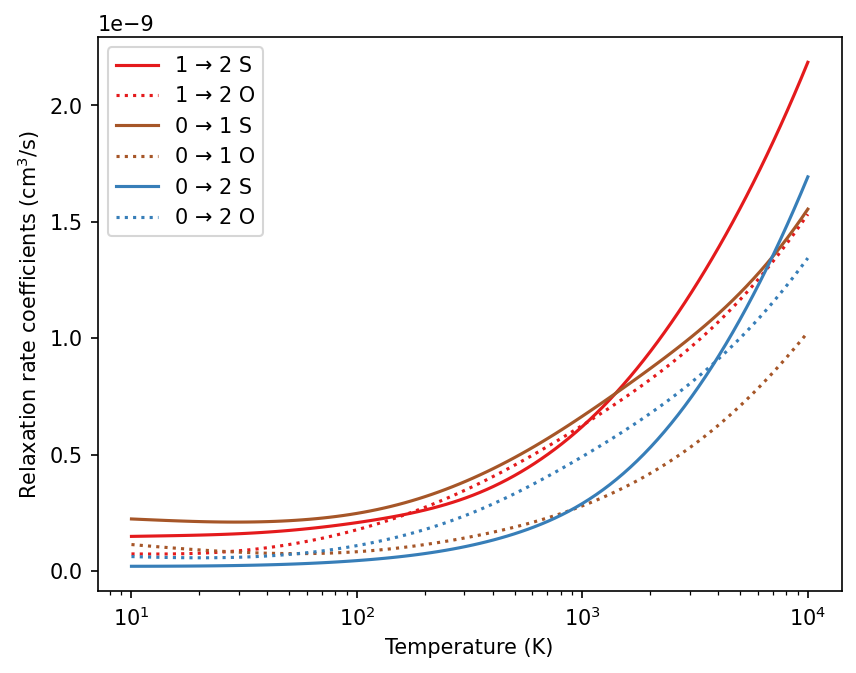}	
    \caption{Excitation and relaxation rate coefficients for fine-structure transitions of S in collision with H (solid lines),
    present work, 
    and for O in collision with H (dotted lines), from   \citet{yan_babb_fine-structure_2022}. The three transitions $(j\rightarrow j')$ are labeled with different colors.}
    \label{fig:SExrate}
\end{figure}

\section{Results}\label{RES}

Rate coefficients as functions of temperature are obtained by averaging the cross sections over a Maxwellian energy distribution,
\begin{equation}\label{rate}
k_{j_a \rightarrow j'_a}(T)=\bigg(\frac{8}{\pi \mu k_B^3T^3}\bigg)^{1/2}\int_0^{\infty} \sigma_{j_a \rightarrow j'_a} (E_k) e^{-E_k/k_{B}T}E_k dE_k \,,
\end{equation}
where $T$ is the temperature, $k_B$ is the Boltzmann constant, and $E_k$ is the kinetic energy. 
The scattering equations were integrated from $R=1.2\,a_0$ to $R=30\,a_0$.
Over this range, accordingly the potentials were utilized as described in Sec.~\ref{pot}
with the potential values for $R<9.83\,a_0$ interpolated using
cubic splines.
Cross sections were calculated using Eq.~(\ref{cross}) from threshold, Eqs.~(\ref{FinE_Si}) and (\ref{FinE_S}), to sufficiently high collisional energies ($10^5$~K or about 8.62~eV) to ensure that the integration in Eq.~(\ref{rate}) could be accurately evaluated up to temperatures of 10,000~K.
[Note that the $\textrm{Si}(^1D)$-$\textrm{H}$ and $\textrm{S}(^1D)$-$\textrm{H}$ channels, respectively, open at about 0.8~eV and 1.2~eV. Our cross sections do not include atomic Si or S electronic excitations higher than the ground $^3P$ terms.]

In Fig.~\ref{fig:SiExsct} we present our calculated excitation and relaxation cross sections of the fine structure transitions for the Si-H system.
The corresponding rate coefficients are given in Fig.~\ref{fig:SiExrate} and in Table~\ref{tab:sihrate}. 
In Fig.~\ref{fig:SiExrate}, we also show the fine-structure excitation and relaxation rate coefficients for the C-H system~\citep{yanbabb2023} for the purpose of comparison. 
Because Si and C are isovalent and share the same fine-structure level ordering with $j_a$ the relaxation cross sections for Si and H collisions, Eq.~(\ref{FinE_Si}), are similar to those for C and H (dotted lines)~\citep{yanbabb2023}, though we find the Si-H values are larger than the C-H values for relatively high energy $T > 1000~\,\mathrm{K}$. 
Also we find that the $0\rightarrow 1$ excitation rate coefficients for Si and H collisions are much larger than those for C and H collisions.

For the S atom in collision with H, the excitation and relaxation cross sections for the fine structure transitions as functions of energy are shown in Fig.~\ref{fig:SExsct}, where we find that the cross sections for relaxation for S and H show some resonances for collisional energies around 10~K. 
The corresponding rate coefficients are given in Fig.~\ref{fig:SExrate} and Table~\ref{tab:shrate}. 
In Fig.~\ref{fig:SExrate}, we also show our calculations on fine-structure excitation and relaxation rate coefficients for the O-H system~\citep{yan_babb_fine-structure_2022} for the purpose of comparison.
We find that the excitation rate coefficients for S and H  are similar to those of O and H.
However, the $0\rightarrow 1$ relaxation rate coefficient for S and H is also much larger than that for O and H.

From the Figs.~\ref{fig:SiExsct} to~\ref{fig:SExrate}, we can find that the cross sections and rate coefficients of Si in collision with H are relatively larger than those of S in collision with H, which are caused by the more "attractive" (deeper) potentials of SiH 
compared to SH (the a$^4\Sigma^-$ electronic state of SiH has numerous bound ro-vibrational levels, while that of SH is repulsive).
The present calculations for Si in collision with H and for S in collision with H can replace rate coefficients estimated by scaling values for Si in collision with He and for S in collision with He values~\citep{Lique-SiHe-SHe-2018}.

\begin{figure}	\includegraphics[width=\columnwidth]{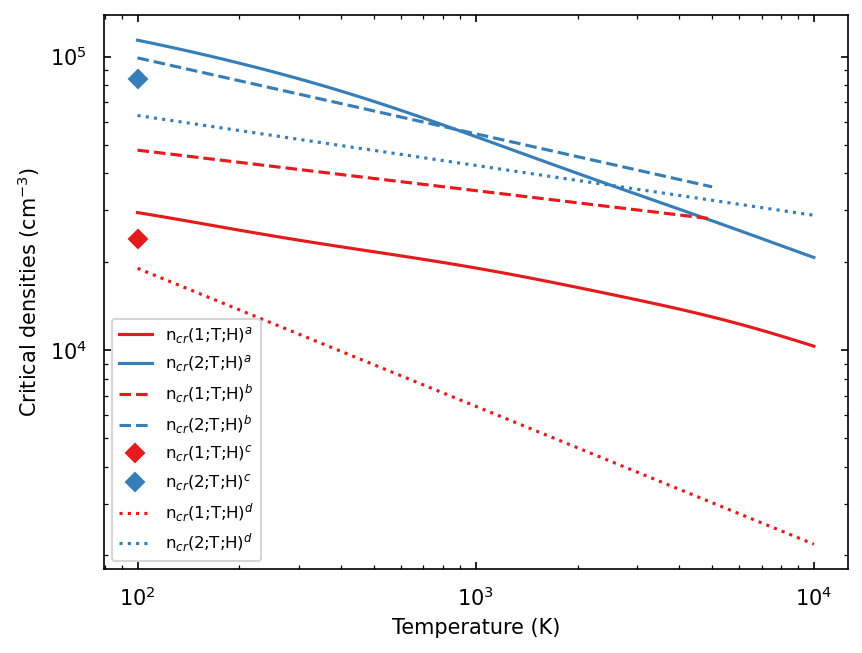}
    \caption{For relaxation collisions of Si with H, comparison of critical densities of the present full quantum close-coupling calculation ($a$ solid lines) with those of \citet[Table 17.1]{Draine2011} ($b$ dashed lines), with those of~\citet[Table 4]{Tielens1985} ($c$ diamond marker) and with those of semiclassical calculation from~\citet[Table 8]{Hollenbach1989} ($d$ dotted lines), where blue denotes the $(2\rightarrow 1)$ transition and red denotes the $(1\rightarrow 0)$ transition. Note that the $(2\rightarrow 0)$ transition is forbidden~\citep{Hollenbach1989}.} 
    \label{fig:Sidensity}
\end{figure}

\begin{figure}	\includegraphics[width=\columnwidth]{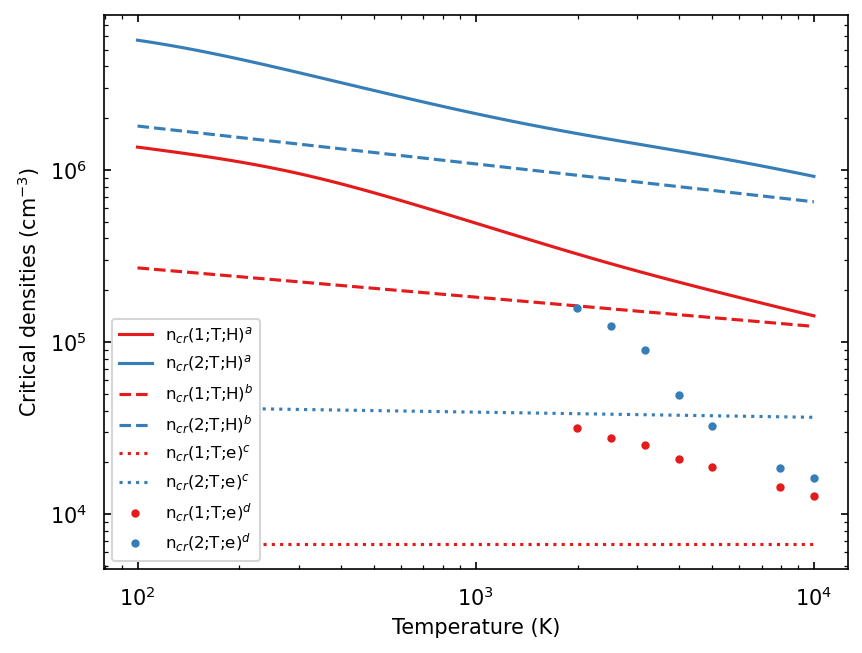}
    \caption{For relaxation collisions of S with H and with electrons, $a$ solid lines are the present full quantum close-coupling calculations for critical densities of the relaxation of S in collision with hydrogen,  $b$ dashed lines are the semiclassical calculations for critical densities of the relaxation in S-H collisions from~\citet[Table 8]{Hollenbach1989}.  For relaxation in S-electron collisions, $c$ small dotted lines are the critical densities from~\citet{Hollenbach1989} and $d$ large dotted lines are the critical densities calculated using the quantum R-matrix calculations of ~\citep{Tayal2004}. 
    The blue denotes the $(1\rightarrow 2)$ transition and red denotes the $(0\rightarrow 1)$ transition. Note that the $(2\rightarrow 0)$ transition is forbidden~\citep{Hollenbach1989}.}
    \label{fig:Sdensity}
\end{figure}

The critical density $n_c$ is defined as
\begin{equation}
n_{c}(j;T;x)
=\frac{\sum_{j'} A(j \rightarrow j')}{\sum_{j'} k_{j \rightarrow j'}(T;x) }\,,
\end{equation}
where $A(j \rightarrow j')$ is the transition probability~\citep{mendoza_recent_1983} and $k_{j \rightarrow j'}(T;x)$ is the relaxation rate coefficient for collisions with species $x$, which maybe be hydrogen atoms (H) or electrons (e).
For the collisions of Si with H, the present calculations of critical densities are shown in Fig.~\ref{fig:Sidensity} along with the values  from \citet[Table 17.1]{Draine2011} and from \citet{Tielens1985}.
For the collisions of S with H, the present calculations are shown in Fig.~\ref{fig:Sdensity} along with the semiclassical calculations of \citet{Hollenbach1989}. 
In Fig.~\ref{fig:Sdensity}, 
we also present the critical densities for the relaxation of S in collisions with electrons which we evaluated using the quantum R-matrix effective collision strengths from~\citet{Tayal2004}; for comparison, we also include the
the semiclassical calculations of \citet{Hollenbach1989}.

\section{Conclusion}\label{Con}

Fine-structure excitation and relaxation cross-sections and rate coefficients for the Si and S atoms in collision with atomic hydrogen are obtained by using quantum-mechanical close-coupling methods. 
The electronic potential curves of SiH and SH are obtained by using the multireference configuration interaction Douglas-Kroll-Hess (MRCI-DKH) method.
The critical densities for the Si and S atoms in H collisions and their comparisons with other calculations or with electron collisions are also given.
The present calculations may be useful in diagnostics of astrophysical environments such as shocked and circumstellar environments.     

\section*{Data Availability}
The rates for collisional relaxation by H presented in Table~\ref{tab:sihrate} for Si and in Table~\ref{tab:shrate} for S are given in LAMDA format~\citep{moldata2005} at Figshare~\citet{YanBabbFig23SiSH}.

\section*{Acknowledgements}
This work was supported by NASA APRA grant 80NSSC19K0698.

\bibliographystyle{mnras}
\bibliography{reference} 
\bsp
\label{lastpage}
\end{document}